# Rejoinder: Demystifying Double Robustness: A Comparison of Alternative Strategies for Estimating a Population Mean from Incomplete Data

**Joseph D. Y. Kang and Joseph L. Schafer**

## 1. CLARIFYING OUR POSITION ON DOUBLY ROBUST ESTIMATORS

We are grateful to the editors for eliciting comments from some of the most prominent researchers in this exciting and rapidly developing field. After we drafted our article, a number of important works on DR estimators appeared, including Tan's (2006) article on causal inference, the monograph by Tsiatis (2006) and the recent articles and technical reports cited by Robins, Sued, Lei-Gomez and Rotnitzky. The discussants' insightful remarks highlight these recent developments and bring us up to date.

Our purpose in writing this article was to provide unfamiliar readers with gentle introduction to DR estimators without the language of influence functions, using only simple concepts from regression analysis and survey inference. We wanted to show that DR estimators come in many different flavors. And, without minimizing the importance of the literature spawned by Robins, Rotnitzky and Zhao (1994), we wanted to draw attention to some older related methods from model-assisted survey sampling which arrive at a similar position from the opposite direction.

*Joseph D. Y. Kang is Research Associate, The Methodology Center, 204 E. Calder Way, Suite 400, State College, Pennsylvania 16801, USA e-mail: josephkang@stat.psu.edu. Joseph L. Schafer is Associate Professor, The Methodology Center, The Pennsylvania State University, 204 E. Calder Way, Suite 400, State College, Pennsylvania 16801, USA e-mail: jls@stat.psu.edu.*

Despite the good performance of $\hat{\mu}_{OLS}$ in our simulated example, we have not and would not argue that it be used routinely and uncritically. The pitfalls of relying solely on outcome regression or $y$-modeling have been well documented for causal inference, where the rates of missing information are high and the impact of selection bias can be dramatic (e.g., Rubin, 2001). Nor do we wish to cast clouds of suspicion over all DR estimators in all circumstances. In many situations, they do work well. On the other hand, we still believe that procedures motivated by parametric $y$-models, when carefully implemented, remain a viable option and should not be categorically dismissed.

Under ignorability, the propensities $\pi_i = P(t_i = 1|x_i)$, $i = 1, \ldots, n$, play no role in likelihood-based or Bayesian inference about $\mu$ under a given $y$-model. If we had absolute faith in one parametric form for $P(y_i|x_i)$, then we could discard all information beyond the sufficient statistics for that model. But the propensities carry information that helps us evaluate the quality of the $y$-model, and we ignore this extra information at our peril, because no model is above criticism. No sensible statistician would argue that propensities should not be examined. But reasonable persons may differ over what role the propensities should play in formulating an estimator. Those who favor a semiparametric approach devise influence functions that combine inverse-propensity weights with regression predictions for $y$. Parametric modelers, on the other hand, may well argue that if the propensities reveal weaknesses in the $y$-model, then that model should be revised and corrected. The latter view has been expressed by Elliott and Little (2000) in the context of survey estimation, where the selection probabilities are known, but parallels to uncontrolled nonresponse and causal inference are obvious.

We believe that propensities are useful for model diagnosis and estimation, but we are still not convinced that they need to enter an influence function as inverse-propensity weights. The strength of weighting is that, if done properly, it protects an estimate from bias regardless of how $y$ is distributed. But this strength can also be a weakness, because such a high level of protection is not always warranted. If the propensities are unrelated to the linear predictors from a good $y$-model, then weighting may be superfluous. If the propensities are poorly estimated or extreme, then combining weights with the regression predictions may do more harm than good. And if the propensities do reveal weaknesses in the $y$-model, inverse-propensity weights are not the only way to correct them.

## 2. RESPONSE TO TSIATIS AND DAVIDIAN

In their illuminating discussion, Tsiatis and Davidian demonstrate that a wide variety of estimators for $\mu$ can be expressed as the solution to an estimating equation based on an influence function. (One possible exception is the class of estimators based on propensity-score matching, which we have not discussed.) Influence functions present interesting results on semiparametric efficiency, but we find them appealing for other reasons as well. First, they show us how to compute a standard error for whatever estimator we choose. Second, they generalize nicely to finite-population sample surveys with complex designs. Regression techniques for complex surveys, as implemented in software packages like SUDAAN (Shah, Barnwell and Biler, 1997), are based on weighted influence functions, so any of the estimators described by Tsiatis and Davidian can be extended to surveys. Third, if we move on to causal inference, we must address the thorny issue of the inestimable partial correlation between the potential outcomes. Any estimator of an average causal effect makes a working assumption about this correlation (e.g., setting it to zero), but a standard error computed from an influence-function sandwich may still perform well when this working correlation is incorrect.

Tsiatis and Davidian mention that our estimator $\hat{\mu}_{\pi\text{-}cov}$, which incorporates propensity-related basis functions into the OLS procedure, is not consistent under $\mathcal{M}_I \cup \mathcal{M}_{II}$ unless the conditional mean of $y_i$ happens to be a linear combination of the particular basis functions for $\pi_i$ used in the OR model. This is certainly true for the usual asymptotic sequence in which the number of basis functions remains fixed as $n \to \infty$. But if we allow the basis to grow with the sample size (e.g., as in a smoothing spline), then it may become DR (Little and An, 2004). Given a large sample, a good data analyst will tend to fit a richer model than with a small sample. If the analyst is allowed to build a rich OR model that corrects for the kind of inadequacies shown in our Figure 4, then the OLS procedure based on the corrected OR model may be as good as any DR procedure.

We like the suggestion by Tsiatis and Davidian of using a hybrid estimator that combines inverse-propensity weighting for cases with moderate propensity and regression predictions for cases with small propensity, an idea echoed by van der Laan and Rubin (2006). As an alternative to a hard threshold $\delta$ at which the change is made, one could opt for a smoother transition by "weighting" each part of the influence function more or less depending on the estimated propensity. We also agree with Tsiatis and Davidian that estimators in the spirit of $\hat{\mu}_{\pi\text{-}cov}$ deserve more consideration even though they are not DR over $\mathcal{M}_I \cup \mathcal{M}_{II}$ in the usual asymptotic sequence. In the simulations of our article, we expressed $m_i$ as a piecewise constant function of $\hat{\pi}_i$ with discontinuities at the sample quintiles of $\hat{\pi}_i$. Another version of $\hat{\mu}_{\pi\text{-}cov}$ that we have found to work well in many situations uses a linear spline in $\hat{\eta}_i = \log(\hat{\pi}_i/(1-\hat{\pi}_i))$ with knots at the quintiles.

## 3. RESPONSE TO TAN

Tan's important work on regression estimators connects the theory of influence functions to ideas of survey regression estimators and the use of control variates in importance sampling. His remarks and propositions are very helpful for understanding the behavior of IPW, OR and DR methods in realistic settings where all the models are incorrect.

We were initially puzzled by several of Tan's points but, upon further consideration, found them to be very insightful. He states that it is more constructive to view DR estimators as efficiency-enhanced versions of IPW than as bias-corrected versions of OR. We find both views helpful for understanding the nature and properties of DR methods. But, as he explains, there are theoretical reasons to expect that his carefully crafted DR estimators may lead to greater improvement over IPW than over a good OR model, because IPW is conservative whereas OR is aggressive.

We are still unsure why Tan states that IPW extrapolates explicitly whereas OR extrapolates implicitly. To us, fitting an OR model to respondents and using that model to predict for nonrespondents is a very obvious kind of extrapolation, especially if the leverage values for some nonrespondents are large relative to those of the respondents. But his points about extrapolation are well taken. All of our methods extrapolate. The assumption of ignorability is itself an extrapolation.

He also points out that estimating an average causal effect is more subtle than simply estimating the mean of each potential outcome and taking the difference. This distinction is important in a semiparametric approach. A semiparametric method that is optimal for estimating two means may not be optimal for estimating the mean difference. Similarly, a method that is optimal for estimating a population average causal effect may not be optimal for estimating the average effect among the treated, or for estimating differences in average causal effects between subpopulations. As parametric assumptions about the OR model are discarded, it becomes important to tailor the estimation procedure to the estimand, which his regression estimators apparently do.

In Tan's simulations, his alternative model in which the analyst sees $X_4 = (Z_3 + Z_4 + 20)^2$ presents an interesting situation where OLS predicts the $y_i$'s for the respondents almost perfectly ($R^2 \approx 0.99$), but the extrapolated linear predictions for the nonrespondents are biased because the unseen true values of $y_i$ turn sharply away from those predictions in the region of low propensity. This is a perfect illustration of how the uncritical use of $\hat{\mu}_{OLS}$ can lead us astray. But in this example, propensity-based diagnostics reveal obvious deficiencies in the linear model. Taking the initial sample of $n = 200$ observations from our article, we fit the linear model to the respondents and a logistic propensity model to all cases given $X_1$, $X_2$, $X_3$, and Tan's alternative version of $X_4$. A plot of the observed residuals from the $y$-model versus the estimated logit-propensities from the $\pi$-model is shown in Figure 1. The loess curve clearly shows that the OLS predictions are biased in the region of high propensity (where it does not really matter) and in the region of low propensity (where it matters very much). If we account for this trend by introducing the squared linear predictor from the logit model $\hat{\eta}_i^2 = (x_i^T \hat{\alpha})^2$ as one more covariate in the $y$-model, the performance of $\hat{\mu}_{OLS}$ greatly improves. Even better performance is obtained with splines, which tend to predict better than ordinary polynomials over the whole range of $\hat{\eta}_i$'s. We created a linear spline basis for $\hat{\eta}_i$ with four knots located at the sample quintiles of $\hat{\eta}_i$. That is, we added the four covariates

$$(1) \quad \begin{aligned} (\hat{\eta}_i - k_1)_+, & \quad (\hat{\eta}_i - k_2)_+, \\ (\hat{\eta}_i - k_3)_+, & \quad (\hat{\eta}_i - k_4)_+ \end{aligned}$$

to the $y$ model, where $(z)_+ = \max(0, z)$ and $k_1, k_2, k_3, k_4$ are the knots. Over 1000 samples, we found that this new version of $\hat{\mu}_{OLS}$ (which, in our article, we would have called $\hat{\mu}_{\pi\text{-}cov}$) performed as well as any of Tan's estimators in the scenario where both models were incorrect. With $n = 200$, we obtained bias = 0.16, % bias = 5.70, RMSE = 2.78 and MAE = 1.78. With $n = 1000$, we obtained bias = 0.30, % bias = 24.6, RMSE = 1.27 and MAE = 0.88. The performance of Tan's regression estimators in these simulations is impressive. The performance of $\hat{\mu}_{OLS}$ is equally impressive if we allow the analyst to make a simple correction to adjust for the $y$-model's obvious lack of fit.

## 4. RESPONSE TO RIDGEWAY AND MCCAFFREY

Ridgeway and McCaffrey correctly observe that, for estimating propensity scores, there are many good alternatives to logistic regression. In addition to their work on the generalized boosted model (GBM), some have been estimating propensities using classification trees (Luellen, Shadish and Clark, 2005) and neural networks (King and Zeng, 2002).

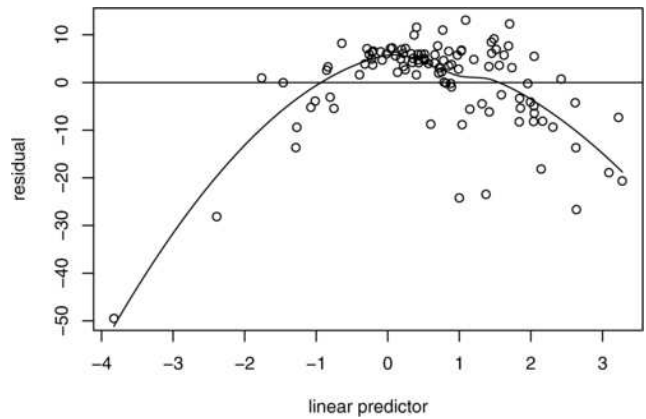

FIG. 1. *Scatterplot of raw residuals from linear y-model fit to respondents in Tan's alternative model, versus the linear predictors from a logistic $\pi$-model, with local polynomial (loess) fit.*

A rich propensity model should improve the performance of the weighted estimator. The advantage of procedures like classification trees and GBM is that they allow us to search through a large space of $\pi$-models, accounting for the effects of many covariates and their interactions, thereby reducing bias in the resulting estimator regardless of how $y_i$ is distributed. These procedures may also reduce variance, because, as explained by Tan, in a sequence of increasingly rich propensity models, the asymptotic variance of an augmented IPW estimator decreases to the semiparametric bound. In principle, one could apply similar procedures like regression trees to create a rich $y$-model. But, as Ridgeway and McCaffrey point out, this raises the possibility of data snooping. As we search through larger and more complicated spaces to find the best $y$-model, it becomes increasingly difficult to compute honest standard errors.

Ridgeway and McCaffrey's simulations with the extra interaction term again reveal the dangers of uncritically relying on $\hat{\mu}_{OLS}$. This interaction increases the degree to which the additive and linear $y$-model is misspecified, so in this scenario we would expect the performance of $\hat{\mu}_{OLS}$ to worsen. The final columns of their Tables 1 and 2 show that, when this interaction is present, propensity-based and DR estimators strongly outperform $\hat{\mu}_{OLS}$. Using the wrong covariates in the propensity model does little harm to the flexible GBM procedure. But one could argue that these comparisons between GBM and $\hat{\mu}_{OLS}$ are unfair in the following sense: They resemble a situation where the analyst is allowed to fit a rich and flexible $\pi$-model but is given no leeway to improve the $y$-model. We examined many samples of $n = 200$ from this new population and found $X_1 X_2$ to be a strong and highly significant predictor of $y$ in every sample. If we add this one interaction to the $y$-model, the bias in $\hat{\mu}_{OLS}$ nearly vanishes, and its RMSE becomes comparable to that of the best DR estimators that Ridgeway and McCaffrey tried. Other interactions are often significant as well. We have not examined the performance of $\hat{\mu}_{OLS}$ when these other interactions are included; doing so would be an interesting exercise.

Our point here is not to argue for the superiority of $\hat{\mu}_{OLS}$ over the DR procedures. Either can work well if applied carefully with appropriate safeguards. And either can be made to fail if we, through the design of a simulation, impose artificial restrictions that force the analyst to ignore clear evidence in the observed data that the procedure is flawed.

## 5. RESPONSE TO ROBINS, SUED, LEI-GOMEZ AND ROTNITZKY

The comments by Robins et al. contain many useful observations and helpful references. Their simulations that reverse the roles of $t_i$ and $1 - t_i$ are instructive. However, in the process of arguing that we misunderstood the message of Bang and Robins (2005), they have apparently misunderstood ours. Their insinuations of cherry-picking might be understandable if we had been arguing for the superiority of $\hat{\mu}_{OLS}$, but that is not what we have done. Quite honestly, we began this investigation fully expecting to demonstrate the *benefits* of dual modeling when neither model is exactly true.

When Bang and Robins (2005) recommended certain DR procedures for routine use, they did so without qualifications or cautionary statements. Now they quote a passage from another article published five years earlier, which Bang and Robins (2005) did not cite, to demonstrate that this was not what they had in mind. Readers cannot react to what they have in mind, but only to what they write. Dr. Robins and his colleagues are eminent researchers, and their statements carry considerable weight. The fact that they knew that these estimators sometimes misbehave but failed to acknowledge it makes their blanket recommendations in 2005 even more troubling.

For the record, we will clarify how we came up with our simulated example. As mentioned in our Section 4, we were trying to loosely mimic a quasi-experiment to assess the average causal effect of dieting on body-mass index among adolescent girls. We decided beforehand that $y_i$ should be predicted from the observed $x_i$ with $R^2 \approx 0.80$, as in the actual data. We decided that the distributions of the estimated propensity scores should resemble those in our Figure 3(e), as in the actual data. We decided that the linear predictors from the $y$-model and $\pi$-model should have a correlation of at least 0.5, as in the actual data, so that $\bar{y}_1 = \sum_i t_i y_i / \sum_i t_i$ would be a strongly biased as an estimator of $\mu$. We decided that the covariates in $x_i$ should not be normally distributed, but they should not be so heavily skewed that a data analyst would need to transform them to reduce the leverage of a few large values. We decided that $x_i$ must be a one-to-one transformation of the unseen true covariates $z_i$ over the effective support of the $z_i$ (without this condition, nonresponse would not be ignorable given $x_i$). Finally, we decided that the linear regression of $y_i$ and the logistic regression of $t_i$ on $x_i$ would be misspecified to about the same extent, in the sense that the correlations between the linear predictors from each model and the corresponding true linear predictors would be about 0.9.

After considerable trial and error, we came up with one example that met all of these criteria. As we ran our simulations, we were truly surprised to see $\hat{\mu}_{OLS}$ perform as well as it did, consistently beating all competitors. We expected that at least some of the DR estimators would improve upon $\hat{\mu}_{OLS}$, but none did. In fact, we were tempted to look for a different example that would demonstrate some of the benefits of DR, but we decided against it precisely because we wanted to *avoid* cherry-picking.

As Robins et al. deconstruct our simulated example, they suggest that our misspecified linear model $E(y_i) = x_i^T \beta$ is so close to being true that $\hat{\mu}_{OLS}$ is virtually guaranteed to outperform all competitors. If that were so, then why did the DR estimators $\hat{\beta}_{WLS}$ and $\hat{\mu}_{BC\text{-}OLS}$ not perform as well, as those estimators were given the same opportunity to take advantage of this nearly correct $y$-model? And, if that were so, why would $\hat{\mu}_{OLS}$ perform so poorly in their simulations when the roles of $t_i$ and $1 - t_i$ were reversed?

The first plot in Figure 1 by Robins et al. reveals that (a) the model for $y_i$ given the vector of true covariates $z_i$ is a linear with very high $R^2$ and (b) the nonresponse is ignorable, so that $P(y_i \mid z_i, t_i = 1)$ and $P(y_i \mid z_i, t_i = 0)$ are the same. This plot implies that conditions where the analyst is allowed to see the $z_i$'s are unrealistic, because knowing $z_i$ is essentially equivalent to knowing $y_i$. But this plot says nothing about the performance of $\hat{\mu}_{OLS}$ or any other estimator when $z_i$ is hidden and the analyst sees only $x_i$, which is the only scenario that we have claimed is realistic. [In fact, the first simulated example published by Bang and Robins (2005) yields a similar picture, because their true data-generating mechanism is also linear and their $R^2$ is 0.94.] The conditional variance $V(y_i \mid z_i)$ was one of many parameters that we had to adjust to create an example that satisfied all of the criteria that we have mentioned. We tried to set $V(y_i \mid z_i)$ to larger values, but doing so decreased the signal-to-noise ratio in the observed data to the point where we no longer saw meaningful biases in any estimators when $n = 200$.

With their Figure 2, Robins et al. purport to show that our misspecified linear regression model fits so well that the predicted values $x_i^T \hat{\beta}$ are essentially unbiased predictions of the missing $y_i$'s, which guarantees excellent performance for $\hat{\mu}_{OLS}$. They state, "We can see that the predicted values of the nonrespondents are reasonably centered around the straight line even for those points with predicted values far from the predicted values of the respondents." On the contrary, our linear model $E(y_i) = x_i^T \beta$ does not give unbiased predictions for nonrespondents or respondents, especially not in the region of extrapolation. To illustrate, we took one simulated sample of $n = 1000$ observations, regressed $y_i$ on $x_i$ among the respondents, and computed the regression predictions $x_i^T \hat{\beta}$ and residuals $y_i - x_i^T \hat{\beta}$ for both groups. A plot of the residuals versus the regression predictions is displayed in Figure 2, along with local polynomial (loess) trends. Respondents are shown in black, and nonrespondents are shown in gray. (For visual clarity, only 20% of the points are displayed, but the loess trends are estimated from the full sample.) For each group, the least-squares regression model strongly underpredicts near the center and overpredicts at the extremes. The reason why $\hat{\mu}_{OLS}$ performs well in this example is not that the linear model is approximately true, but that *the positive and negative residuals in the nonrespondent group approximately cancel out.* The average value of $y_i - x_i^T \hat{\beta}$ for respondents is exactly zero (a consequence of OLS), and the average value of $y_i - x_i^T \hat{\beta}$ for nonrespondents is close to zero. Over 1000 simulated samples, the average of $y_i - x_i^T \hat{\beta}$ among nonrespondents was 1.68. Multiplying this by $-0.5$ (because the average nonresponse rate is 50%) gives $-0.84$, the estimated bias for $\hat{\mu}_{OLS}$ reported in our Table 3.

Figure 2 also reveals why $\hat{\mu}_{OLS}$ was not beaten in this example by any of the dual-modeling methods. The differences between the two loess curves in Figure 2 are not large, showing that the OLS predictions have similar patterns of bias for respondents and nonrespondents. When the predictions from a $y$-model are biased, and the biases are similar when $t_i = 1$ and $t_i = 0$, they are not easily corrected by an estimated propensity model.

If we reverse the roles of $t_1$ and $1 - t_i$, as Robins et al. have done, the situation dramatically changes. Taking the same sample of $n = 1000$, we regressed $y_i$ on $x_i$ when $t_i = 0$ and predicted the responses for both groups. Residuals versus predicted values from this reverse fit are shown in Figure 3. (Once again, for visual clarity, only 20% of the sampled points are shown, but the loess trends are estimated from the full sample.) For the $t_i = 0$ group, the linear model underpredicts at the center and overpredicts at the extremes, and the average value $y_i = x_i^T \hat{\beta}$ is zero. But for the $t_i = 1$ group, the linear model consistently overpredicts across the entire range, introducing a strong upward bias into $\hat{\mu}_{OLS}$.

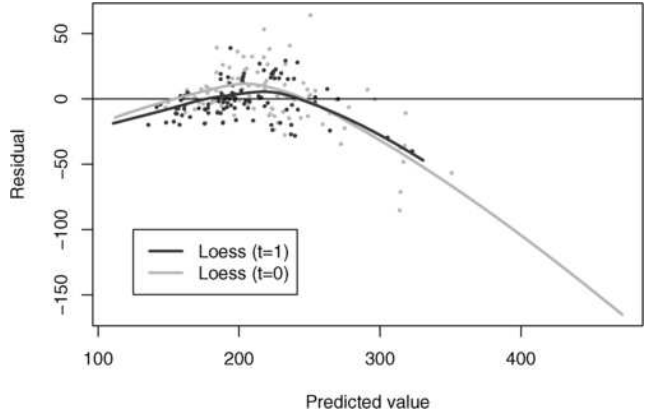

FIG. 2. *Residuals versus predicted values for respondents ($t_i = 1$) (black dots) and nonrespondents ($t_i = 0$) (gray dots) from one sample of $n = 1000$ from our original simulation, with local polynomial (loess) trends for each group. For visual clarity, only 20% of the sampled points are shown.*

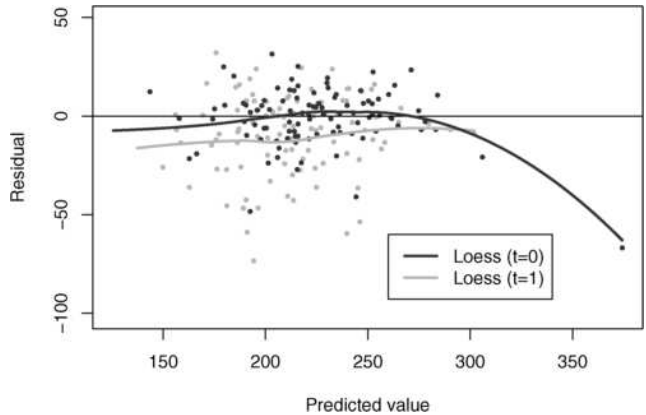

FIG. 3. *Plot analogous to Figure 2, with the roles of $t_i$ and $1 - t_i$ reversed. Cases with $t_i = 0$ and $t_i = 1$ are denoted by black and gray dots, respectively, with local polynomial (loess) trends shown for each group. For visual clarity, only 20% of the sampled points are shown.*

This alternative simulation by Robins et al. is a classic example where patterns of bias in a linear $y$-model cause $\hat{\mu}_{OLS}$ to perform poorly. But because the patterns are dramatically different when $t_i = 0$ and $t_i = 1$, it is also a classic example where the failure can be readily diagnosed and corrected by fitting a $\pi$-model. A plot of the residuals $y_i - x_i^T \hat{\beta}$ for the $t_i = 0$ group versus the linear predictors from a logistic propensity model is shown in Figure 4. The plot, which is based only on $(x_i, t_i, (1 - t_i)y_i)$, shows a strong tendency for the linear $y$-model to overpredict when $P(t_i = 1)$ is low or high. To correct this bias, we created a spline basis as in expression (1), with knots at the sample quintiles, and included the four extra terms as predictors in the linear $y$-model. The performance of $\hat{\mu}_{OLS}$ (which we would now call $\hat{\mu}_{\pi\text{-}cov}$) improved dramatically, and the new estimator worked better than any of the dual-modeling methods reported by Robins et al. The performance statistics in the both-models-wrong scenario were Bias = 2.21, Var. = 12.61 and MSE = 17.46 when $n = 200$, and Bias = 2.40, Var. = 1.88, and MSE = 7.66 when $n = 1000$, which compare favorably to the results shown by Robins et al. in their Table 2.

## 6. CONCLUDING REMARKS

As statisticians devise newer and fancier methods, we hope to find one that is foolproof, yielding good results no matter when and how it is applied. But the search for a foolproof method is quixotic and futile. Some procedures are, on balance, better than others, but each one requires many subjective inputs, and none should be applied routinely or uncritically. As we develop better estimators, we should also strive to give potential users a healthy dose of intuition about how the procedures work, their limitations, sound recommendations about their use, and diagnostics that can help users decide when a procedure is trustworthy and when it is not.

In conclusion, we believe that propensity modeling is prudent and even necessary when rates of missing information are high. But we are still not convinced

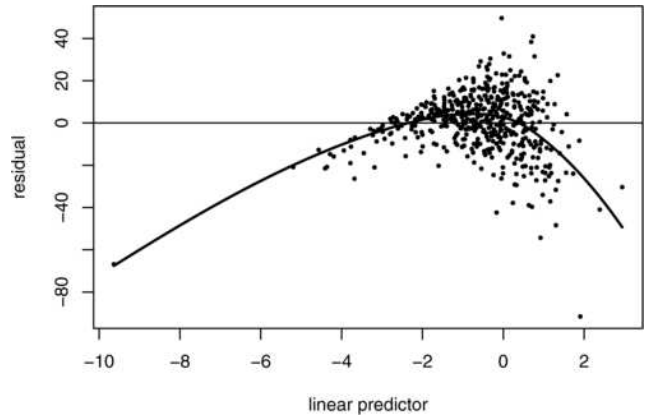

FIG. 4. *Scatterplot of residuals from a linear $y$-model fit to $t_i = 0$ cases, versus linear predictors from a logistic $\pi$-model, with local polynomial (loess) fit in one sample of $n = 1000$ from the alternative simulation study by Robins et al.*

that estimated inverse propensities must always be used as weights.

## ACKNOWLEDGMENT

This research was supported by National Institute on Drug Abuse Grant P50-DA10075.